\documentstyle[12pt,epsf]{article}

\newcommand{\beq}{\begin{equation}}
\newcommand{\eeq}{\end{equation}}
\newcommand{\bea}{\begin{eqnarray}}
\newcommand{\eea}{\end{eqnarray}}

\newcommand{\ba}{\begin{array}}
\newcommand{\ea}{\end{array}}
\newcommand{\be}{\begin{equation}}
\newcommand{\ee}{\end{equation}}
\newcommand{\beas}{\begin{eqnarray*}}
\newcommand{\eeas}{\end{eqnarray*}}

\newcommand{\nbox}{{\,\lower0.9pt\vbox{\hrule \hbox{\vrule height 0.2 cm \hskip
0.2 cm \vrule height 0.2 cm}\hrule}\,}}

\DeclareFixedFont{\xiiss}{OT1}{cmss}{m}{n}{12}
\DeclareFixedFont{\ixss}{OT1}{cmss}{m}{n}{9}
\DeclareFixedFont{\cmrnine}{OT1}{cmr}{m}{n}{9}

\newcommand{\Tr}{{\rm Tr}\ }
\newcommand{\STr}{{\rm STr}\ }
\newcommand{\CC}{\hbox{\xiiss C\kern-.4emI}}
\newcommand{\RR}{\hbox{\xiiss R\kern-.45emI}}
\newcommand{\ZZ}{\hbox{\xiiss Z\kern-.4emZ}}
\newcommand{\CCs}{\hbox{\ixss C\kern-.4emI}}
\newcommand{\ZZs}{\hbox{\ixss Z\kern-.4emZ}}
\newcommand{\pa}{\partial}

\newcommand{\pasl}{\pa\kern-.55em /}
\def\href#1#2{#2}

\begin{document}
\begin{titlepage}

\begin{flushright}
SU-ITP 00-32\\
RU-NHETC-2000-46\\
hep-th/0012065
\end{flushright}

\vskip 1cm

\begin{center} {\Large \bf

Gauge Invariant Correlators in

\vspace{4mm}

Non-Commutative Gauge Theory} 
 
\end{center}

\vspace{1ex}

\begin{center}
{\large
Moshe Rozali$^a$ and Mark Van Raamsdonk$^b$}

\vspace{5mm}
${}^a${ \sl Department of Physics and Astronomy} \\
{ \sl Rutgers University} \\
{ \sl Piscataway, NJ 08855} \\
{ \tt rozali@physics.rutgers.edu}

\vspace{5mm}
${}^b${ \sl Department of Physics} \\
{ \sl Stanford University} \\
{ \sl Stanford, CA 94305 U.S.A.} \\
{ \tt mav@itp.stanford.edu}

\end{center}

\vspace{2.5ex}
\bigskip
\centerline {\bf Abstract}

\bigskip

Using perturbation theory, we explore the universal high momentum
behavior of correlation functions of gauge invariant operators in
planar noncommutative gauge theories. We find that the correlation
functions are strongly enhanced when pairs of momenta become antiparallel. In 
particular, there is a transition from the previously noted exponential 
suppression of correlation functions at high momenta to a more field theoretic
behavior when the momenta of pairs of operators antialign within a
critical angle.  Some of our calculations can be extrapolated to  strong
coupling, and in particular we are able to reproduce precisely the 
supergravity prediction for the behavior of two point functions, including
the coupling dependence.

\bigskip

\end{titlepage}

\newpage

\section{Introduction}

Many of the recent interesting developments in string theory have
centered around various non-gravitational theories that arise from
string theory in ``decoupling'' limits. These include local quantum
field theories, such as the ${\cal N}=4$ SYM theory, plus non-local theories
without a known explicit description, such as little string theories. 
Somewhere in between are the non-commutative
field theories which arise in certain decoupling limits involving
D-branes with NS-NS two-form fields turned on in their worldvolume
directions \cite{cds,dh,schomerus,sw}. These theories are non-local, yet possess 
explicit Lagrangian
descriptions in the language of ordinary quantum field theory that
allow perturbative calculations.

Perturbative analysis of noncommutative field theories has revealed
many interesting properties (see e.g. \cite{msv,sus}). In particular, the
theories 
display a
mixing of the infrared and the ultraviolet: short distance
physics can lead to significant long range effects. From a technical point of
view, these effects arise from internal momentum dependent phase
factors which appear in non-planar graphs of the theory. On the other
hand, planar graphs behave identically to their counterparts in the
commutative theory apart from overall external momentum dependent
phase factors. One might then expect that large N noncommutative field
theories should have basically the same behavior as their commutative
counterparts, since non-planar diagrams are suppressed by powers of
${1 \over N}$. However, it turns out that at least for noncommutative
gauge theories, the commutative and noncommutative versions have very different
behavior at large momenta, even in the planar, large $N$ limit.

This differing behavior can be seen easily for the particular case of
${\cal N}=4$ SYM theory by comparing the conjectured supergravity duals for
the two cases. In the commutative case, the corresponding
gravitational theory is of course type IIB string theory on $AdS^5 \times S^5$
which arises as the near horizon geometry of a stack of 
D3-branes \cite{maldacena}. In
the noncommutative case, the dual theory is type IIB string theory in
a background which has been worked out in
 \cite{hi,mr}. The two solutions have identical behavior in the ``infrared''
part of the space (near $r=0$), suggesting that the two field theories
are equivalent at low energies.  On the other hand, the supergravity dual of 
the noncommutative theory has an Einstein frame metric \cite{drt} which is
{\it asymptotically flat} in the $r \to \infty$ region (though the coupling
goes to zero here) in sharp contrast with the asymptotically AdS dual
of the commutative theory. Thus, we expect that the UV behavior of
noncommutative ${\cal N}= 4$ SYM theory differs  
significantly from the behavior of
its commutative counterpart, even for large $N$. The
different behavior in the ultraviolet reflects the fact that unlike
conventional field theories, non-commuatative gauge theories are not
defined by a UV fixed point \cite{micha}. It seems particularly
interesting to understand this UV behavior, since it provides the field
theory dual of gravity on a space which is not asymptotically AdS.

In this paper, we study the high momentum behavior of correlation functions
of gauge invariant operators in noncommutative gauge theory. As discussed in
\cite{iikk}-\cite{dw} and reviewed in section 2 of this paper,
these operators involve open
Wilson lines which extend in the directions perpendicular to the operator 
momentum with a length proportional to the momentum. This transverse 
spreading at high momenta is in accord with the uncertainty relation 
associated 
with the commutation relations for the coordinates, $[x^\mu, x^\nu] = i 
\theta^{\mu \nu}$. The extended nature of the gauge invariant 
operators leads to 
qualitatively different behavior of high momentum correlation functions as 
compared with the commutative theory.

In \cite{ghi}, the two point functions of these operators were calculated at weak coupling and shown to have a universal exponential behavior with high momentum,
\be
\label{twoweak}
\langle O(k) O(-k) \rangle \sim {\rm exp} \left( \sqrt{{g^2 N |k||k \theta| \over 4
\pi}} + \dots \right)
\ee
where the dots indicate a regularization dependent term that cancels out in 
properly normalized correlation functions. The authors showed that the 
momentum dependence of this expression is identical to that of the two point function calculated using the supergravity dual of noncommutative ${\cal N} = 4$ SYM theory, though the supergravity result has a different coupling dependence, obtained by the replacement $g^2 N \to \sqrt{g^2 N}$ in (\ref{twoweak}). 
In this paper, we show that even this coupling dependence can be reproduced by a perturbative field theory calculation extrapolated to strong coupling. We describe this result in section 4 and offer two possible explanations for the agreement.

In the rest of the paper we consider the high momentum behavior of 
more general correlation functions. We find a rather 
universal effect: the correlation functions are very sharply enhanced when the 
momenta of pairs of operators becomes antiparallel.\footnote{This is
consistent with the interpretation of elementary quanta in the theory
as dipoles, advanced in \cite{bs}.} In particular, for
a pair of operators with large nearly antiparallel large momenta of 
order $k$,  ($|k||\theta k| \gg 1$) we find that the correlation functions 
vanish rapidly when the angle 
between the momenta rises above some critical value, $\phi > {1 \over |k| 
|\theta k|}$. As a result of this enhancement amplitudes fall off
exponentially
with the momenta for generic angles, but cross-over to a more field
theoretic behavior below this  critical small angle.

The structure of the paper is as follows. In section 2, we review the 
construction of gauge invariant operators in noncommutative gauge theories and 
discuss in particular the operators corresponding to supergravity modes in the 
gauge theory - gravity correspondence for ${\cal N} = 4$ SYM
theory. In section 
3, we provide a general formula for computing planar correlation functions of 
operators involving open Wilson lines. In section 4, we discuss the 
calculation 
of two-point functions and in section 5  we extend this discussion to
higher point functions. We offer some comments and a summary of the results 
in section 6. 

\section{Gauge Invariant Operators}

In this section, we review the construction of gauge invariant operators in  
noncommutative gauge theories as well as the specific form of
operators corresponding to bulk supergravity modes in the gauge theory
- gravity correspondence for noncommutative ${\cal N} =4$ SYM theory. 

To begin, we recall the transformation properties of the gauge field
in noncommutative gauge theory, 
\be
A_\mu \rightarrow U(x) \star A_\mu \star U(x)^\dagger -{i \over g}
U(x) \star \partial_\mu U^\dagger (x)
\ee
We will take a $U(N)$ gauge group, so that $A_\mu$ is an $N \times N$ hermitian
matrix. The gauge covariant field strength is then given by
\be
F_{\mu \nu} = \partial_\mu A_\nu - \partial_\nu A_\mu + ig(A_\mu \star
A_\nu - A_\nu \star A_\mu)
\ee
and the (Euclidean) gauge invariant action for the gauge fields is
\be
S = {1 \over 4} \int d^4 x \Tr(F_{\mu \nu} \star F_{\mu \nu}) 
\ee
(equivalently, the star product may be replaced by an ordinary
product). In more general theories such as the ${\cal N} = 4$ SYM
theory, we may have additional matter fields, but the generalization of
the action to a noncommutative theory is always obtained by replacing
 products by star products.

We now turn to the construction of gauge invariant operators. A
remarkable property of noncommutative gauge theories (with all fields
transforming in the adjoint) is that translations in the
noncommutative directions are gauge transformations,
\be
\label{prop}
f(x^\mu + \theta^{\mu \nu} k_\nu) = e^{i k \cdot x} \star f(x^\mu) \star e^{-ik 
\cdot x}  
\ee
As a result, it is impossible to construct local gauge invariant
operators. On the other hand, it is possible to construct gauge
invariant operators that are local in momentum space, as shown in 
\cite{iikk,ghi,dasrey,dw}.

\begin{figure}
\centerline{\epsfysize=1.5truein \epsfbox{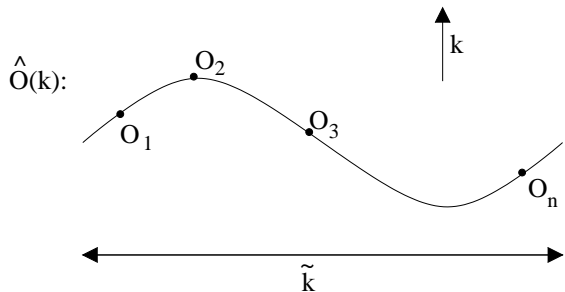}}
 \caption{A general Wilson line operator with momentum $k$.}
\label{line}
\end{figure}

The construction employs open Wilson lines, defined by
\bea
W(x, C) &=& \Tr \left( {\rm P}_\star e^{ig \int_C A_\mu dx^\mu}
\right) \nonumber\\
&=& \sum_{n=0}^\infty (ig)^n \int_0^1 d \sigma_1 
\cdots \int_{\sigma_{n-1}}^1 d
\sigma_n \dot{\zeta^{\mu_1}} \cdots \dot{\zeta^{\mu_n}} \nonumber\\
&& \hspace{1.3in} \Tr \left( A_{\mu_1}(x+
\zeta(\sigma_1)) \star \cdots \star A_{\mu_n}(x + \zeta(\sigma_n))
\right)
\nonumber \\
\eea
where $P_\star$ denotes path ordering and $C$ is a path parameterized by $\zeta(\sigma)$ with $\zeta(0) = x$. These 
transform covariantly under gauge transformations,
\be
W(x,C) \rightarrow U(x) \star W(x,C) \star U^\dagger(x + \zeta(1))
\ee
Now, given a set of gauge covariant local operators $\{ {\cal O}_i \}$
transforming in the adjoint, we may construct a  gauge invariant
operator local in momentum space by taking
\be
\hat{\cal O}(k) = \int e^{i k \cdot x} \Tr( {\cal O}_1 (x) \star W(x,C_1)
 \star {\cal O}_2 (x + l_1) \star \cdots \star W(x +
l_1 +\dots + l_{n-1} ,C_n) )      
\ee
where $l_i = \zeta_i(1)$. Using the relation (\ref{prop}), it is
simple to show that this operator is gauge invariant provided that 
\be
l^\mu_1 + \dots + l^\mu_n = \theta^{\mu \nu} k_\nu
\ee
This operator is precisely the result of taking insertions
of any set of local covariant operators at arbitrary point on an
open Wilson line whose endpoints are separated by $\tilde{k} \equiv 
 \theta^{\mu \nu} k_\nu$, as diagrammed in figure \ref{line}.

In fact, it is somewhat misleading to refer to these objects as open
Wilson lines, as the endpoints of the interval are not
distinguished. Noting 
that
\be
\int e^{i k \cdot x} \,\Tr( {\cal O}_1 (x) 
\star W(x,C) \star {\cal O}_2 (x + 
\tilde{k})) = \int e^{i k \cdot x}\, \Tr({\cal O}_2(x) \star {\cal O}_1(x) \star 
W(x,C)) \; ,
\ee
we see that the operators at each end of the open Wilson line may equivalently 
be taken at a single point by a cyclic rearrangement. 

For the case of noncommutative ${\cal N} = 4$ SYM theory, there is now 
compelling evidence \cite{ghi,hong,triv} that the operators corresponding to particle
states in the dual gravitational theory are gauge invariant operators
whose form is a special case of the construction described so
far. Recall that for the usual commutative ${\cal N} = 4$ theory,
operators corresponding to particles in $AdS^5 \times S^5$ are chiral
operators which take the form of a symmetrized trace of a product of
gauge covariant objects (scalars, fermions, field strengths, or
covariant derivatives of these),
\be
{\cal O}(k) = \int e^{i k \cdot x}\, \STr(B_1 \cdots B_n)(x)
\ee
It turns out that the appropriate generalization of this operator to the 
noncommutative theory is given by 
\be
{\cal O}(k) = \int e^{i k \cdot x}\, {\rm \hat{STr}}(B_1 \star \cdots \star B_n 
\star W(x,\tilde{k}) )
\ee
where $W(x, \tilde{k})$ is a Wilson line with a straight line path
from $x$ to $x+\tilde{k}$ and ${\rm \hat{STr}}$ denotes that the
expression is to be averaged over all ways of inserting the $B$s
into separate points on the the Wilson line. Thus, the individual
factors in the product become spread out linearly over a transverse
distance $\tilde{k}$.


In the following sections we  study general correlation functions
of these straight Wilson line operators, in particular looking for 
universal behavior at high momenta that arises from the extended
nature of 
the operators.
It is interesting to note that due to the asymptotically flat nature of the 
supergravity dual (in the maximally supersymmetric case),  there are 
scattering states in the dual geometry, and it is likely that the 
correlations functions we compute  correspond to S-matrix elements of 
these scattering states. This is similar to the situation in linear
dilaton backgrounds \cite{abks}, and is in contrast to the
asymptotically AdS geometries.

\section{Correlation Functions of Gauge Invariant Operators}

We have seen that the gauge invariant operators in noncommutative gauge theories 
have an extended structure that differs significantly from the corresponding 
operators in the commutative theory at high momenta. In this section, we 
investigate the effects of this extended structure on general correlation 
functions.

Since we are mainly interested in the effects of the Wilson line part of the 
operators, we focus on the simple case of Wilson lines with a single insertion 
of a gauge covariant operator ${\cal O}$. We define
\bea
W(k) &=& \int d^4 x \; e^{i k \cdot x} \star {\rm Tr} \left(
{\cal O}(x) \star {\rm P}_\star e^{ig \int A_\mu dx^\mu} \right) \nonumber\\
&=& \sum_{n=0}^\infty (ig)^n \int d^4 x \; e^{i k \cdot x}  
\int_{0 \le \sigma_1 \le \cdots \le \sigma_n \le 1}  \dot{
\zeta}^{\mu_1}(\sigma_1) \cdots \dot{\zeta}^{\mu_n}
(\sigma_n) \nonumber \\
&& \hspace{0.5in}  {\rm Tr} \left( {\cal O}(x) \star A_{\mu_1} (x +
\zeta(\sigma_1)) \star \cdots \star A_{\mu_n} (x + \zeta(\sigma_n)) \right)
\label{opexp}
\eea
As before, ${\rm P}_\star$ denotes noncommutative path ordering, and 
the Wilson line  runs over a path $x^{\mu} = \zeta^\mu(\sigma)$
, with $\zeta^{\mu}(1) - \zeta^{\mu}(0) = \tilde{k}^{\mu} \equiv
\Theta^{\mu \nu} k_\nu$. We will mainly consider the case of a
straight Wilson line for which $\zeta^\mu (\sigma) = \tilde{k}^{\mu}
\sigma$.   

We focus on the large $N$ theory, for which planar diagrams are
dominant. For such  diagrams, the star products appearing in the
expansion (\ref{opexp}) only have the effect of providing an overall phase
factor for the correlation function, which depends on the ordering of
the external momenta. 

\begin{figure}
\centerline{\epsfysize=1.5truein \epsfbox{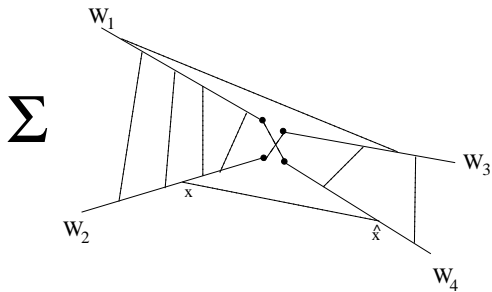}}
 \caption{A typical planar ladder diagram in the correlation function
of four Wilson line operators }
\label{feyn}
\end{figure}

To focus on the UV behavior of the theory, we assume that the external momenta 
are asymptotically large. More precisely, we assume that the dimensionless 
quantity $|k||\tilde{k}| $ is large. The amplitudes are 
then dominated at each order in the 't Hooft coupling by ladder diagrams,
as discussed in \cite{ghi}. These are diagrams, such as the one 
shown in figure \ref{feyn}, in which the gauge fields are
contracted between various Wilson lines, with no internal 
vertices.\footnote{Briefly, the dominance of ladder diagrams at large momenta 
arises because the endpoints of each ``rung'' are integrated over the (long) 
Wilson lines, giving rise to factors of $\tilde{k}$. These factors
enhance the ladder diagram relative to any other diagram of  the same
order in  the 't Hooft coupling.} Since we also work at leading order in the 
't Hooft $\frac{1}{N}$ expansion, we are further restricted to planar ladder diagrams.

For planar diagrams in which the Wilson line gauge
fields contract only with each other, it is straightforward write down
a general expression for the correlation function. We find
\bea
\langle W(k_1) \cdots W(k_n) \rangle_{\rm ladder} &=&  \sum e^{i \sum_{i<j}
 k_i  \times k_j}  \int d^4 x_i \;
e^{i k_i \cdot x_i}  \langle {\cal O}_1 (x_1) \cdots {\cal O}_n (x_n)
\rangle \nonumber \\
&& \hspace{0.6in} \times \int d \vec{\sigma}  \prod {g^2 N
\over 4 \pi^2} {- \dot{\zeta} \cdot \dot{\hat{\zeta}}
 \over (x - \hat{x})^2}
\label{general}
\eea
Here, the first sum is over all ways of contracting the $A$s between the various Wilson lines
to give a planar diagram. The phase factor comes from the star
products, but depends only on the ordering of external momenta since
the diagrams are planar. In the second line, the integral runs over
positions of the $A$s on the various lines (i.e. the endpoints of the
propagators), and the product is over all propagators
in the diagram. In the integrand, $x$ and $\hat{x}$ represent the two endpoints 
of a given
propagator, while $\zeta$ and $\hat{\zeta}$ are tangent vectors evaluated
at $x$ and $\hat{x}$ respectively. A typical diagram in the sum is shown in 
figure \ref{feyn}.

\section{Two Point Functions}

We start by calculating two point functions of the gauge invariant 
operators (\ref{opexp}). The calculation was performed previously for weak coupling in \cite{ghi}. Here, we note that within the approximation of summing only over ladder diagrams, the result may also be
evaluated at strong coupling. We find that this strong coupling result exactly reproduces the behavior of the supergravity result including the coupling dependence. At the end of this section we offer two possible explanations for the agreement. A similar result was obtained in \cite{esz} in the context of Wilson line correlators for the commutative ${\cal N}=4 $ theory, see also \cite{dg}. 

For the two point functions, factoring out the $\delta$-function of momentum 
conservation, the general formula (\ref{general}) becomes
\beq
\left< W(k)W^\dagger(k)\right> = \int d^4 \Delta \,  e^{ik\cdot
\Delta} 
f(\Delta)
\sum_n (\lambda \tilde{k}^2)^n \prod_{i=1}^n \int d\tau_i \int d\sigma_i \,
\frac{1}{|\Delta + \tilde{k}(\tau_i - \sigma_i)|^2}
\eeq

Here $\lambda = {g^2 N \over 4 \pi^2}$, $\Delta = x_1 -x_2$ is the relative
 position of the operators, and we define 
\beq
\left< O_1(x_1) O_2(x_2) \right> = f(x_1 - x_2) \; .
\eeq
Because of the restriction to planar ladder diagrams, the integrals over 
$\sigma$ and $\tau$ (which describe the endpoints of the various rungs) are ordered, $0 \le \sigma_1 \le \cdots \le \sigma_n \le 1$, 
$0 \le \tau_1 \le \cdots \le \tau_n \le 1$.
In order to extract the universal large momentum behavior arising from the 
Wilson lines, we follow \cite{ghi} and set $f(\Delta)=1$.

The integrand $\frac{1}{|\Delta + \tilde{k}(\tau_i - \sigma_i)|^2}$ is a
 steep function (for large $|\tilde{k}|$)  which is maximized at:
\beq
\tau_i - \sigma_i  =  - \frac{\Delta_\parallel}{|\tilde{k}|}
\eeq
where $\Delta_\parallel$ is the component of $\Delta$ parallel to $\tilde{k}$.
In order to attain this maximum within the integration region one requires:
\beq
\label{Drange}
0 \leq \Delta_\parallel \leq |\tilde{k}|
\eeq

Since the integrand is steep, we would like to evaluate it using a saddle point 
approximation starting with the $\tau_i$ integrals. However, due to the ordering 
of the points, $\tau_i \leq \tau_{i+1}$, the peak of the integrand may be close 
to the edges of the integration regime, so the saddle point approximation is not 
automatically justified. The required condition is that the width of the 
integrand as a function of a given $\tau_i$, 
\[
\frac{|\Delta_\perp|}{|\tilde{k}|}
\]
is less than the average spacing between the $\tau_i$'s, ${1 \over n}$. Thus, to 
use the saddle point approximation, we need
\beq
\eta \equiv \frac{n |\Delta_\perp|}{|\tilde{k}|} \leq 1
\eeq
When this condition is satisfied, defining
\beq
I_n = \prod_{i=1}^n \int d\tau_i \int d\sigma_i 
\frac{\lambda \tilde{k}^2}{|\Delta + \tilde{k}(\tau_i - \sigma_i)|^2} \; ,
\eeq
we find (first evaluating the $\tau$ integrals using saddle point and then 
performing the $\sigma$ integrals directly),
\beq
\label{saddle}
I_n \sim \frac{\lambda^n}{n!}\,\frac{\tilde{k}^n}{|\Delta_\perp|^n}
\eeq
To see that this estimate  cannot hold when $\eta \ge 1$, note that the integral 
$I_n$ is 
bounded from above by the maximum value of the integrand times the volume of the 
integration region
\beq
\label{estimate}
I_n \leq \frac{\lambda^n}{(n!)^2}\,\frac{\tilde{k}^{2n}}{\Delta_\perp^{2n}}
\eeq
It is easy to see that the saddle point result (\ref{saddle}) exceeds 
this bound 
when $\eta > 1 $. In fact, the upper bound (\ref{estimate}) provides a good 
approximation to $I_n$ for $\eta \ge 1$ (we will show this explicitly
in section 5 using a lower bound on $I_n$).

In either regime, the integral over $\Delta_\parallel$ (which we restrict to the range (\ref{Drange})) can be readily performed, remembering that 
$k_\parallel =0$. Suppressing the subscript of $\Delta_\perp$, we get:
\beq
\left< W(k)W^\dagger(k)\right> \simeq |\tilde{k}|\int d^3\Delta
\, e^{ik \cdot \Delta}
\sum_n (\frac{\lambda \tilde{k}^2}{4\pi})^n I_n
\eeq
where 
\be
I_n \sim \left\{ \ba{ll} {\lambda^n \over n!} \left( {\tilde{k} \over \Delta} \right)^n & \qquad \qquad {n \Delta \over \tilde{k}} < 1 \\
 {\lambda^{n} \over (n!)^2} \left({\tilde{k} \over \Delta }\right)^{2n} & \qquad \qquad {n \Delta \over \tilde{k}} > 1
\ea \right.
\ee

The correct estimate for $I_n$ depends now on the values of $\Delta, n$ 
dominating the expression for the two-point function. For both estimates
 of $I_n$, the integration over $\Delta$ is dominated by the regime $|\Delta| 
\simeq  \frac{n}{|k|}$, therefore one has $\eta \simeq \frac{n^2}{|k||\tilde{k}|}$ in the relevant region of integration. Thus, for $n < \sqrt{k \tilde{k}}$ we may use the $\eta < 1$ estimate when integrating over $\Delta$, while for $n > \sqrt{k \tilde{k}}$, we should use the $\eta >1$ estimate.

As explained in \cite{ghi}, we should extract the contribution which is non-analytic in $k$ for each $n$, since the analytic parts correspond to contact terms in position space. The results for the relevant Fourier transforms can be found in \cite{ghi}. We are left with a sum of terms $\sum_n J_n$ where the $n$ dependence of $J_n$ is given by 
\be
J_n \sim \left\{ \ba{ll} {\lambda^n (k \tilde{k})^n \over (n!)^2} & \qquad \qquad n < \sqrt{k \tilde{k}} \\
 {\lambda^n (k \tilde{k})^{2n} \over (n!)^2 (2n)!} & \qquad \qquad n > \sqrt{k \tilde{k}}\ea \right.
\label{Jrange}
\ee
For each value of the parameters, the terms in this series will
increase to a maximum value at some $n=M$ and then decrease, with the
terms for $n \gg M$ giving negligible contribution. At weak coupling,
we find that $M \ll \sqrt{k \tilde{k}}$ so only the small $n$ form of
$J_n$ is relevant. Performing the sum using the small $n$ estimate, we recover precisely the result (\ref{twoweak}) of \cite{ghi}. 

On the other hand, at strong coupling, $M \gg \sqrt{k \tilde{k}}$ so the series is dominated by terms of the large $n$ form. 
%
Summing the series in this case, we find (ignoring prefactors)
\beq
\left< W(k)W^\dagger(k)\right> \sim  \exp(\sqrt{\sqrt{\lambda} 
|\tilde{k}| |k|})
\eeq

We note that up to numerical factors (which we have not been careful
about), this result is precisely equal to the supergravity result \cite{ghi}
including the coupling dependence. 

We offer two possible explanations for this agreement. Firstly, in the case
of chiral operators for the ${\cal N}=4$ theory, it is reasonable to guess 
that two point functions should obey a non-renormalization theorem
such that the leading perturbative result is valid at all values of
the coupling. But in our case, the leading
perturbative result is precisely this sum over ladder diagrams. The
higher order ladder diagrams are not perturbative corrections but
actually the leading contribution to the correlation function from the
higher order terms in the expansion of the operators (\ref{opexp}).
Thus, assuming a nonrenormalization theorem, we would expect that the
sum over ladder
diagrams at strong coupling should reproduce the supergravity answer. This
is exactly what we find, so our result may be interpreted as providing 
evidence for this nonrenormalization theorem. 

A second possibility for the agreement is that as noted earlier, at high momenta the ladder diagrams give the leading contribution order by order in the 't Hooft coupling, while diagrams with internal vertices are suppressed by inverse powers of momenta. It is therefore possible that in the large momentum limit, the restriction to a sum over ladder diagrams is sensible even at strong coupling, providing another possible explanation of the agreement we find. If this is the case, it would indicate that both chiral and nonchiral operators in the theory share this universal behavior at large momenta. From the supergravity point of view, this would indicate a similar behavior in the UV region for both supergravity modes and stringy modes. 

\section{Higher Point Functions}

 We turn now to higher point functions of the noncommutative operators. We will show that for large momenta, the correlation functions are sharply peaked, preferring the kinematics in which pairs of momenta are antiparallel. The properly normalized correlation functions generically exhibit exponential decay at large  momenta, as described in \cite{ghi}. However, at small enough angles we find that the amplitudes are enhanced, and the dependence on momenta has a cross-over to a more typical field theoretic behavior, as described below.

We consider then the correlation functions $\langle W_1(k_1) W_2(k_2)  \cdots 
W_n(k_n)\rangle$. In the limit of large $k_i$ in the sense described above, we 
will show that ladder diagrams must generally be included even at weak 
coupling, as was the case for the two point functions. As we have discussed,   
the ladder diagrams dominate any diagrams with internal vertices at each order in the 't Hooft coupling. Diagrams with internal vertices are suppressed by powers of ${1\over |k||\tilde{k}|}$, so working in the order of limits in which the momenta are taken large first, we may consider only the ladder diagrams.
These may be summed explicitly both for weak and strong coupling, and we provide the results for both cases. We emphasise that summing over ladder diagrams at strong coupling can be justified only with our carefully chosen order of limits.   The agreement with the gravity results for the two-point function is an indication that the sum over ladder diagrams may be sensible even in the limit of strong coupling.

In subsection 5.1 we derive the conditions under which it is necessary to  
include ladder diagrams connecting a given pair of operators $W_1, W_2$.
We will see that in a weak coupling calculation and for generic angles between 
the momenta $\tilde{k_i}$, there is no need to include ladder diagrams, and we may use the 
leading perturbative result. However, when two momenta $\tilde{k_1},\tilde{k_2}$ 
are nearly anti-parallel, the sum over ladders connecting $W_1, W_2$
is necessary, and provides an enhancement of the amplitude. At strong coupling, ladder diagrams will be important for all angles.

In section 5.2 we present the approximation schemes which are appropriate for 
evaluation of this universal behavior in both the weak and strong coupling 
regimes. These approximation schemes generalize the two estimates ($\eta < 1$ 
and $\eta >1$) for the two point functions, presented in the previous section. In the sections 5.3 and 5.4, we perform the calculation of the ladder diagram contributions in each of these regimes.

The reader interested only in the results of the calculation is invited to skip 
to section 6, where our results are summarized and discussed.

\subsection{Conditions for Including Ladder Diagrams}

Given a general correlation function, we would first like to determine the 
conditions under which it is necessary to include ladder diagrams connecting a 
pair of operators $W_1(k)$ and $W_2(-l)$ in a perturbative calculation.\footnote{ We have 
written the momentum of $W_2$ as $-l$ because this will simplify the equations.}  We will find that ladder diagrams dominate the leading order perturbative result when the angle between $\tilde{k}$ and $\tilde {l}$ is less than a particular 
upper bound.

To begin, consider a diagram involving an additional
contraction between gauge fields in the Wilson lines associated with $W_1$
and $W_2$. Relative to the leading order perturbative result (with no 
contractions between the Wilson lines) this diagram will have an additional 
factor of
\be
\lambda (\tilde{k} \cdot \tilde{l}) \,  \int_0^1 d \sigma \int_0^1
d \tau {1 \over (\Delta + \tilde{k} \sigma - \tilde{l} \tau)^2}
\ee
We wish to determine when this factor is of order 1 or larger, in which case the 
diagram (and higher order ladder diagrams) must be included.

For simplicity, we assume that $\Delta$ is perpendicular to $\tilde{k}$ and 
$\tilde{l}$, which are taken to have the same magnitude. In this case, the 
integral may be evaluated explicitly, and the relative factor becomes
\be
\lambda  \cot(\phi) \sinh^{-1} \left( {\tilde{k} \sin {\phi
\over 2} \over \Delta} \right) \; ,
\ee
where $\phi$ denotes the angle between $\tilde{k}$ and $\tilde{l}$, assumed to 
be small. The distance $\Delta$ lies purely in the commutative
directions.

The typical value of $\Delta$ for operators of momentum $k$ will be
roughly $\Delta = {1 \over |k|}$, so we estimate that the ladder diagram
contributions will be important for 
\be
{\lambda \over \phi} \sinh^{-1} ( |k| |\tilde{k}| \phi) > 1
\ee

\begin{figure}
\centerline{\epsfysize=2truein \epsfbox{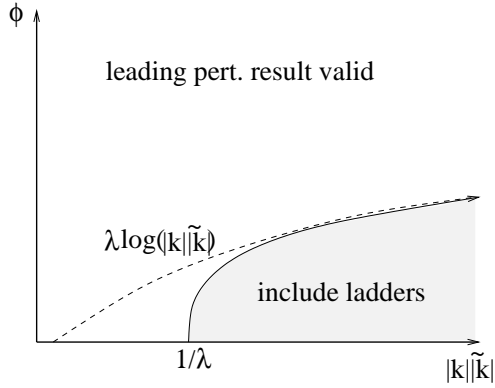}}
 \caption{Angles and momenta for which ladder diagrams dominate the
leading perturbative result.}
\end{figure}

This region is diagrammed in figure 3. We see that ladder diagrams must be 
included for small angles whenever $|k||\tilde{k}| \ge 1/\lambda$. In 
particular, for momenta such that $|k||\tilde{k}| \gg 1/\lambda$, ladder 
diagrams must be included when
\be
\phi < \lambda \log(|k||\tilde{k}|) \; .
\ee
At larger angles, the ladder diagrams do not make a significant contribution, 
and we can trust the leading order perturbative result. At weak coupling, this 
cutoff angle remains small ($\ll 1$) as long as $|k||\tilde{k}| < e^{1 \over 
\lambda}$, so the leading order perturbative result is fine for generic angles 
$(\phi \sim 1)$. On the other hand, it is obvious that ladder diagrams will dominate 
the leading perturbative result at strong coupling for any angles.  

\subsection{Approximation Schemes}

We would like to calculate the contribution to the correlation function
\be
\langle W_1(k) W_2(-l)  \cdots W_n \rangle 
\ee
from the sum over ladders connecting the Wilson lines in $W_1,W_2$ in the case 
where $k$ and $-l$ are very large and nearly antiparallel.
We are interested in universal behavior arising from the extended nature of the 
Wilson lines, independent of the details of the operators $W_i$.

\begin{figure}
\centerline{\epsfysize=1.5truein \epsfbox{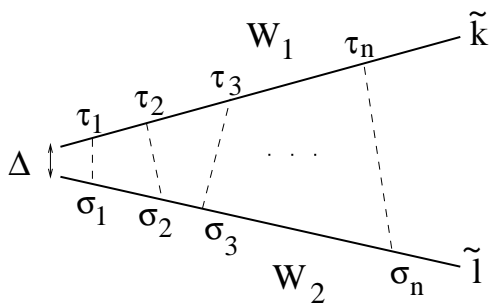}}
 \caption{Simple configuration of Wilson lines.}
\label{sym}
\end{figure}

From the general formula (\ref{general}), we see that the correlation function 
has the following dependence on $k, l$
\be
\label{expr}
\langle W_1(k) W_2(-l)  \cdots W_n \rangle \sim \int d^4 \Delta 
\exp(i\Delta \cdot (k - l)/2 ) f(\Delta, \cdots) \sum_n {\cal I}_n 
\ee
where $f$ represents the commutative position space correlator, and ${\cal I}_n$ 
is the contribution coming from the ladder diagram with $n$ rungs connecting 
$W_1$ and $W_2$ (see figure \ref{sym}), 
\beq
\label{integral}
{\cal I}_n = \int_0^1 d \sigma_n \cdots \int_0^{\sigma_2} d \sigma_1
\; \int_0^1 d \tau_n \cdots \int_0^{\tau_2} d \tau_1 \; 
(\tilde{k} \cdot \tilde{l})^n
\; \prod_{i=1}^n{ \lambda  \over |\Delta + \tilde{k} \sigma_i - \tilde{l} 
\tau_i|^2 }
\eeq

We would like to determine the behavior of the integral ${\cal I}_n$ for large 
$\tilde{k}$ and $-\tilde{l}$ equal in magnitude (for simplicity) and nearly 
antiparallel. In the full expression for the momentum space correlator, the 
separation
$\Delta$ is integrated over, so we are mainly
interested in the behavior of the integral in the region of $\Delta$ where
the integral is largest.  

\begin{figure}
\centerline{\epsfysize=3truein \epsfbox{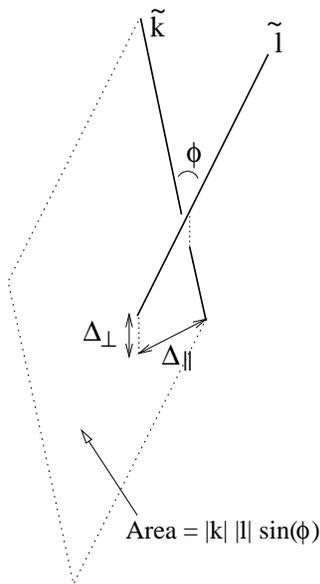}}
 \caption{Maximally contributing configurations of two Wilson lines.}
\label{area}
\end{figure}

Let $\Delta_\perp$ and $\Delta_\parallel$ be the components of $\Delta$
perpendicular to and parallel to the plane of noncommutativity.\footnote{For unitarity, we assume that the noncommutativity parameter $\Theta^{\mu \nu}$ is nonzero only for a pair of spatial directions and refer to these directions as the plane of noncommutativity.} The largest
values of ${\cal I}_n$  occur if
$\Delta_\parallel$ is such that the two lines cross in a projection onto the 
plane of noncommutativity. From figure \ref{area} it is easy to
 see that this region of
$\Delta_\parallel$ has area $|\tilde{k}| |\tilde{l}| \sin(\phi)$, where $\phi$ 
is
the angle between $\tilde{k}$ and $\tilde{l}$. 

For $\Delta_\parallel$ in the region described, the minimum distance
between the two lines will be $|\Delta_\perp|$. We expect that the main
contribution to the integral will be from the region in which most of the
propagators have length of the same order of magnitude as this minimum
length. In this case, most of the propagator endpoints remain within
a distance of roughly ${|\Delta_\perp| \over \phi}$ from the 
crossing point. The
average separation between endpoints on each line will therefore be roughly
$\min \left({|\Delta_\perp| \over n \phi},  {|\tilde{k}| \over
n}\right)$. 

If the average separation between the endpoints is much greater than
the average propagator length, it is a good approximation to extend the
region of integration for each $\tau$ between $-\infty$ and $\infty$. This
approximation will therefore be valid if 
\be
\label{bound}
n < \min({1 \over \phi}, {|\tilde{k}| \over |\Delta_\perp|}) 
\ee

We will call this the small $n$ approximation, since there is an upper
bound on $n$ (though it may be large). This generalizes the $\eta < 1$ regime in 
our evaluation of two point function for which the saddle point approximation 
was justified (\ref{saddle}). For $n$ larger than this bound, we will require 
other methods to evaluate the integral. 

We now turn to a detailed evaluation of the integral ${\cal I}_n$ in 
both the small and large $n$ approximations. It will turn out 
that for weak coupling calculations, only the small $n$ regime is relevant, 
while for strong coupling, we require the large $n$ behavior.

\subsection{Small $n$ Approximation}

We begin by evaluating the integral ${\cal I}_n$ in the small $n$ regime. As 
discussed, for $n$ satisfying the bound (\ref{bound}) it is a good
approximation to extend the integration region for the $\tau$'s to the
whole real line for a typical set of $\sigma$'s (equivalent to a saddle point 
approximation). This also provides an
upper bound on the integral for all values of $n$ and the other parameters.

After performing this saddle point evaluation, the remaining integral takes the 
form
\bea
{\cal I}_n = \int_0^1 d \sigma_n \cdots \int_0^{\sigma_2} d 
\sigma_1 f(\sigma_1) \cdots
f(\sigma_n) &=& {1 \over n!} (\int_0^1 d \sigma f(\sigma))^n \nonumber\\
&\equiv& {1 \over n!} H^n
\eea
where 
\bea
f(\sigma) &=& \int_{-\infty}^\infty d \tau {\lambda \tilde{k} \cdot \tilde{l}
\over | \Delta + \tilde{k} \sigma - \tilde{l} \tau|^2 } \nonumber \\
&=& {\lambda \tilde{k} \cdot \tilde{l} \over  \sqrt{ \tilde{l}^2 (\Delta +
\tilde{k} \sigma)^2 - (\tilde{l} \cdot (\Delta + \tilde{k} \sigma))^2}}
\eea
We can perform the the integral over $\sigma$ exactly, though the result is
somewhat messy. In order to get an idea of the dependence on $\phi$ and
$\tilde{k}$, it is illuminating to consider the case in which the lines
have equal length. 

The maximum contribution comes from the value of $\Delta_\parallel$
for which the lines cross symmetrically in the projection to the 
noncommutative directions.
This occurs for $\Delta \cdot \tilde{k} = - {\tilde{k}^2 \sin^2 \phi /2}$,
and in this case, the result is 
\be
H = \lambda \cot \phi \, \log \left
( {\sqrt{({\tilde{k} \sin \phi \over 2 \Delta})^2 + 1} + {\tilde{k} \sin
\phi \over 2 \Delta} \over  \sqrt{({\tilde{k} \sin \phi \over 2 \Delta})^2
+ 1} - {\tilde{k} \sin \phi \over 2 \Delta} } \right) 
\ee
Here, once again for the sake of brevity we denote by $\Delta$ the 
distance in the commutative directions, previously denoted as $\Delta_\perp$.
Though we have chosen a particular value for $\Delta_\parallel$, we note that ${\cal I}_n$ should be roughly constant over the region of $\Delta_\parallel$ shown in figure 5 since the dominant configurations in the integral will have all propagators clustered around the crossing point.

Recalling that ${\cal I}_n = H^n/n!$, we see that for angles $\phi \ll {\Delta 
\over \tilde{k}}$, 
\be
{\cal I}_n = {1 \over n!} \left(\lambda \, {\tilde{k} \over \Delta} (1 
- {\tilde{k}^2 \phi^2 \over
24 \Delta^2} + \cdots) \right)^n
\ee
while for $\phi \gg {\Delta \over \tilde{k}}$, we get 
\be
{\cal I}_n = {1 \over n!} \left( \lambda \, {1 \over \phi} \ln 
\left( \tilde{k} \phi
\over \Delta \right) \right)^n
\ee
These formulae accurately describe the behavior of ${\cal I}_n$ in the small $n$ 
regime, (\ref{bound}).

\subsection{Large $n$ Approximation}

We now consider the case in which the small $n$ approximation is no longer
valid. This occurs when the average separation between points on each line becomes less than the typical distance between the lines. In this limit, the
propagator endpoints become dense on the two lines, and we expect that most
of the contribution to the integral comes from a small region about some
preferred configuration. 

To get an idea of the $\tilde{k}$ and $\phi$ dependence of ${\cal I}_n$ in
the large $n$ regime, we again take the simple configuration with
$|\tilde{k}| = |\tilde{l}|$ and $\Delta_{\parallel} = 0$ (this is  the 
configuration shown in figure \ref{sym} if the separation $\Delta$ in the diagram is 
understood to be perpendicular to the lines). Since the integrals are difficult 
to evaluate directly in the large $n$ regime, we will determine the behavior of ${\cal I}_n$ by establishing upper and lower bounds.

\subsubsection{Large $n$, Upper Bound}

First, we will determine an upper bound on ${\cal I}_n$ for this
configuration using the relation 
\be
\int |f_1 \cdot f_2 \cdots f_n| \le (\int |f_1|^n)^{1 \over n} \cdots
(\int |f_n|^n)^{1 \over n} 
\ee
where $f_i$ are arbitrary functions on some integration region (this
is a generalization of the Cauchy-Schwarz inequality). This implies that
\be
\label{upper}
{\cal I}_n \le (\prod_{m=1}^n \hat{{\cal I}}_m)^{1 \over n}
\ee
where $\hat{{\cal I}}_m$ is obtained from ${\cal I}_n$ by replacing the product 
of
propagators in the integrand by $n$ copies of the $m$th propagator. The
integrals $\hat{{\cal I}}_m$ are much easier to evaluate, as we will see. 

The integrand of $\hat{{\cal I}}_m$ depends only on $\sigma_m$ and $\tau_m$, and
it is straightforward to perform the integrals over all other $\sigma$s and
$\tau$s. We find
\be
\hat{\cal I}_m =  (\lambda \tilde{k} \cdot \tilde{l})^n \int_0^1 d \sigma_m 
\int_0^1
 d \tau_m  \; {f(\sigma_m) \; f(\tau_m) \over |\Delta + \tilde{k} \sigma_m
 - \tilde{l} \tau_m|^{2n}}
\ee
where
\be
f(t) = {t^{m-1} (1-t)^{n-m} \over (m-1)! (n-m)!}
\ee
We note that $f(t)$ is sharply peaked at $t = {m-1 \over n-1}$ while the
denominator of $\hat{\cal I}_m$ is sharply peaked near
$\sigma=\tau=0$. The integral will receive most of its contribution
from these two regions. For small enough $\Delta$, it is the latter
region near the origin that dominates, since, as we will see, ${\cal
I}_n$ diverges as $\Delta \to 0$. It is convenient to rewrite the
integral in terms of $\sigma_+ = \sigma + \tau$ and $\sigma_- 
= \sigma - \tau$. This gives
\bea
\hat{\cal I}_m &=& {\lambda^n \tilde{k}^{2n}\cos^n \phi \over [(m-1)!(n-m)!]^2}\\
&& \hspace{0.3in} {1 \over 2} \int_0^2 d \sigma_+ \int d \sigma_- {({1 \over 4}
 (\sigma_+^2- \sigma_-^2))^{m-1}((1- {1 \over 2} \sigma_+)^2 - {1 \over 4}
 \sigma_-^2)^{n-m} \over (\Delta^2 + \tilde{k}^2 \cos^2({\phi \over 2})
\sigma_-^2 + \tilde{k}^2 \sin^2({\phi \over 2}) \sigma_+^2)^n } \nonumber\\
\eea
The $\sigma_-$ integral  is sharply peaked at $\sigma_-=0$ and may be 
evaluated using the saddle point approximation. Near $\sigma_+ = 0$, the 
result (for $m>1$) is
\be
\label{resgen}
\hat{\cal I}_m = {\lambda^n \tilde{k}^{2n} \cos^n \phi \over 2^{2m-1} [(m-1)!(n-
m)!]^2} 
 \sqrt{ \pi \over m-1}  \int_0 d \sigma_+ { \sigma_+^{2m-1} 
(1- {1 \over 2} \sigma_+)^{2n-2m} \over (\Delta^2 + \tilde{k}^2 \sin^2({\phi 
\over 2})\sigma_+^2)^n}
\ee
The dominant term for small $\Delta$ ($\Delta < \tilde{k} \phi$) may now be 
computed exactly, and we find (for $1 < m < n$)
\be
\hat{\cal I}_m = {\lambda^n \over [(n-1)!]^2} {n-1 \choose m-1} {1 \over 4^m (n-
m) 
\sqrt{m-1} } {\cos^n \phi \over \sin^{2m} {\phi \over 2}} \left( {\tilde{k}^2
 \over \Delta^2} \right)^{n-m}
\ee
Finally, using the relation (\ref{upper}) we find an upper bound for $\phi > 
\tilde{k} / \Delta$  (ignoring constants)
\be
{\cal I}_n \le {\lambda^n \over (n!)^2} \left({\tilde{k} \cos \phi \over \Delta
\sin{\phi \over 2} }\right)^n
\ee
Proceeding from equation (\ref{resgen}) for $\phi < \tilde{k} / \Delta$ we find 
\be
{\cal I}_n \le {\lambda^n \over (n!)^2} \left( {\tilde{k} \over \Delta} 
\right)^{2n} \left(1 - {\tilde{k}^2 \phi^2 \over 3 \Delta^2} + \dots \right)^n
\ee

\subsubsection{Large $n$, Lower Bound}

We now determine a lower bound on ${\cal I}_n$ in the large $n$
regime. To do this, we restrict the integration region by choosing an ordered 
set $0=a_0 \le a_1 \le \cdots \le a_n-1 \le a_n = 1$ such that 
\be
a_i < \sigma_i, \tau_i < a_{i+1} 
\ee
Then certainly,
\be
\label{lb}
{\cal I}_n \ge \prod \int_{a_i}^{a_{i+1}} d \sigma_i
\int_{a_i}^{a_{i+1}} d \tau_i {\lambda \tilde{k} \cdot \tilde{l} \over |\Delta + 
\tilde{k} \sigma - \tilde{l} \tau|^2}\\
\ee
This bound is valid for all choices of  $a_i$, so in
order to establish the best lower bound, we would like to maximize the
last expression over all such choices.\footnote{Note that we have restricted the 
$\sigma$ and $\tau$ integrals in the same way because we are still assuming a 
symmetric configuration in which $\tilde{k}$ and $\tilde{l}$ have the same 
length with $\Delta_\parallel=0$. In more general cases, it would be necessary 
to divide the $\sigma$ and $\tau$ integrals differently to obtain the best 
bound.} Despite a great reduction in the size of 
the integration region, this approach actually gives a reasonable estimate of 
${\cal I}_n$ for large $n$ since the full integral is dominated by contributions 
from a very small region about some preferred configuration, and the 
maximization over choices of $a_i$  essentially selects this dominant 
region. 

Since we take $n$ to be very large, the values $a_i$ are very closely spaced. We 
therefore can define a continuous functions $f(x)$, such that:
\be
a_i = f(\frac{i}{n}) 
\ee
This satisfies the boundary conditions:
\be
\label{boundary}
f(0)=0 \qquad \qquad f(1) = 1
\ee
Taking logarithms of both sides of (\ref{lb}) and replacing the sum by an 
integral we then obtain 
\be
\log \left( {{\cal I}_n \over \lambda^n (\tilde{k} \cdot \tilde{l})^n} \right) 
\ge G_n(f) \; ,
\ee
where
\bea
G_n(f) &=&  n \,\int_0^1 dx \{  2\log(f'(x){1 \over n})
 - \log(|\Delta + \tilde{k} f(x) - \tilde{l} f(x)|^2 )\} \\
&=& -2n\log(n) +n \,\int_0^1 dx \{  2\log(f'(x)- \log({\Delta_\perp}^2 + f^2 
S^2)\} \;
\eea
Here, we have defined $S= 2 |\tilde{k}|\sin(\frac{\phi}{2})$, and to avoid clutter, we omit the subscript in $\Delta_\perp$.

We now maximize $G$ by solving the Euler-Lagrange equation for $f$.
This gives
\beq
2\log(f') -  \log(\Delta^2 + f^2 S^2) = C \\
\eeq
where $C$  is a  constant. 

Solving this and choosing integration
constants to satisfy the boundary conditions (\ref{boundary}), we 
find\footnote{Note: $\sinh^{-1}(x) = \log(x+ \sqrt{1+x^2})$}
\be
f_{min}(x) = {\Delta \over  S} \sinh(x \cdot \sinh^{-1}({ S \over
\Delta}))
\ee
It is now simple to plug this value of $f_{min}$ into $G$ to obtain a lower
bound
\bea
{\cal I}_n &\ge& (\lambda \tilde{k} \cdot \tilde{l})^n e^{G(f_{min})}
\nonumber\\
&=& {\lambda^n \cos^n \phi \over n^{2n}}\left( {1 \over 2 \sin{\phi \over 2}}
\sinh^{-1} \left({2 \tilde{k} \sin{\phi \over 2} \over \Delta }\right) 
\right)^{2n} 
\eea
For $\tilde{k} \phi \ll \Delta$ , which includes the 
exactly antiparallel case,
this gives
\beq
{\cal I}_n \ge {\lambda^n \cos^n \phi \over n^{2n}} \left( {\tilde{k} \over 
\Delta} \right)^{2n} \left( 1 - {1 \over 6} \left( {2 
\tilde{k} \sin {\phi \over 2} \over \Delta} \right)^2 + \dots \right)^{2n}
\eeq
while for $\tilde{k} \phi \gg \Delta$ (the non-parallel case for small
enough $\Delta$) we get
\be
{\cal I}_n \ge {\lambda^n \cos^n \phi \over n^{2n}}\left({1 \over 2 \tilde{k} 
\sin{\phi \over 2}}
\log \left({4 \tilde{k} \sin {\phi \over 2} \over \Delta} \right) \right)^{2n}
\ee
Note also that 
\be
n^{2n} = e^{2(n \ln(n) - n)} e^{2n} \approx
(n!)^2 e^{2n}
\ee
so for example in the antiparallel case, we have upper and lower
bounds
\be
{1 \over (n!)^2} {1 \over (e \Delta)^{2n}} \le {\cal I}_n \le {1 \over
(n!)^2} {1 \over \Delta^{2n}}
\ee

This justifies the claim in section 2 that the upper bound provides a 
good estimate of $I_n$ in the large $n$ regime for the antiparallel case.

We have now determined the behavior of ${\cal I}_n$ for general values
of $n$ and the other parameters. In the next section we summarize these results and discuss the
physical consequences for the behavior of the correlation functions at
various values of the parameters.    

\section{Summary of Results}

We have considered correlation functions of the form
\be
\langle W_1(k) W_2(-l)  \cdots W_n(k_n) \rangle 
\ee
and noted that even at weak coupling, ladder diagrams connecting $W_1$
and $W_2$ must be included when the momenta $k$ and $-l$ are nearly
antiparallel, as summarized in figure 3. For the ladder diagram
contributions, the dependence on $k$ and $l$ is given by
\be
\langle W_1(k) W_2(-l)  \cdots W_n \rangle \sim \int d^4 \Delta 
\exp(i\Delta \cdot (k - l)/2 ) f(\Delta, \cdots) \sum_n {\cal I}_n 
\ee
where $f$ represents the commutative position space correlator and
${\cal I}_n$ is defined in (\ref{integral}).

\begin{figure}
\centerline{\epsfysize=2truein \epsfbox{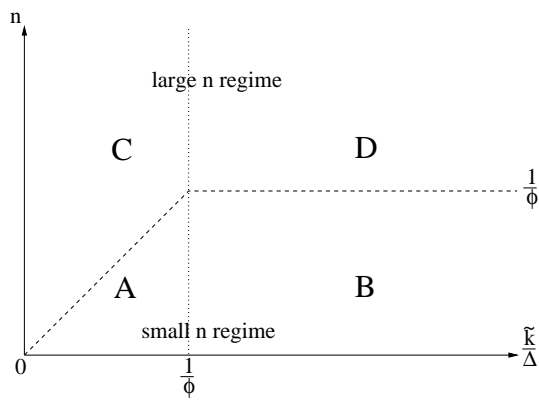}}
 \caption{Regimes of behavior for ${\cal I}_n$ in the $\Delta-n$ plane.}
\label{regimes}
\end{figure}

Our results for the behavior of ${\cal I}_n$ are summarized in figure
\ref{regimes}.  For momenta of magnitude $k$ with a given fixed angle $\phi$ between the, we find four regions of behavior as $n$ and $\Delta$ are varied:\\
\\
{\bf small n:}\\
\\
In this regime, defined by
\be
n < \min({1 \over \phi},{\tilde{k} \over \Delta}) \; ,
\ee
the typical spacing between points on the Wilson lines is greater than the 
average separation between the lines (propagator length). We find that 
\be
{\cal I}_n = \left\{ \ba{ll} {1 \over n!} \left(\lambda \, {\tilde{k} 
\over \Delta}(1 - {\tilde{k}^2 \phi^2 \over 24 \Delta^2} + \cdots )\right)^n  
& \qquad \qquad {\tilde{k} \over \Delta} \ll {1 \over \phi} \\
{1 \over n!} \left(\lambda \, {1 \over \phi} \log({\tilde{k} 
\phi \over 
\Delta}) \right)^n & \qquad \qquad {\tilde{k} \over \Delta} \gg {1
\over \phi}
\ea \right.
 \ee
These correspond to regions $A$ and $B$ respectively in figure \ref{regimes}.\\ 
\\
{\bf large n:}\\
\\
In this regime, for which 
\be
n > \min({1 \over \phi},{\tilde{k} \over \Delta}) \; ,
\ee
the points are densely packed on the two Wilson lines. For ${\tilde{k} \over 
\Delta} > {1 \over \phi}$ (region $D$) we have determined upper and lower bounds given by
\be
{\lambda^n \cos^n \phi \over (n!)^2} \left( {1 \over 2 \sin {\phi \over 2} } 
\log \left( {4 \tilde{k} \sin {\phi \over 2} \over \Delta} \right)
\right)^{2n} \le {\cal I}_n 
\le {\lambda^n \over (n!)^2} \left( { \tilde{k} \cos \phi \over \Delta \sin 
{\phi \over 
2} } \right)^n
\ee
Finally, in region $C$ with ${\tilde{k} \over \Delta} < {1 \over \phi}$ we 
find
\beas
& &{\lambda^n \over n^{2n}} \left({\tilde{k} \over \Delta} 
\right)^{2n}  
\left( 1 - {1 \over 6} \left( {\tilde{k} \phi \over \Delta} \right)^2
+ \dots \right)^{2n} \le {\cal
I}_n \le {\lambda^n \over (n!)^2} \left( {\tilde{k} \over \Delta} \right)^{2n}
\left(1 - {\tilde{k}^2 \phi^2 \over 3 \Delta^2} + \dots \right)^n
\eeas

In evaluating the momentum space correlators, we must sum over $n$ and 
integrate over $\Delta$ as in equation (\ref{expr}). In this case, because of the additional $\Delta$ dependence of the function $f(\Delta)$ in (\ref{expr}), it is simplest to consider first the sum over $n$ with everything else fixed.

Note that as a function of $n$, the series of terms ${\cal I}_n$ is exponential 
(${\cal I}_n \sim {H^n \over n!}$) in the small $n$ regime  and then falls of 
faster (${\cal I}_n \sim {G^n \over (n!)^2}$)in the large $n$ regime. Thus, it is 
obvious that the series converges for any values of the parameters. Furthermore, just as for the two point function, there will be some $M$ such that 
the terms start decreasing for $n > M$, and terms with $n \gg M$ will give negligible 
contribution. 

At weak coupling ($\lambda \ll 1$), we find that $M$ lies well within the small 
$n$ regime as long as\footnote{For ${\tilde{k} \over \Delta} < 
e^{1 \over \lambda}$, this will be true at all angles at which ladder diagrams are important.} 
\be
\phi < {\Delta \over \tilde{k}} e^{1 \over \lambda} \; .
\ee
In this case, the large $n$ terms have negligible contribution and we find
\bea
\ba{ll}
\sum{\cal I}_n = e^{\lambda \,{\tilde{k} \over \Delta}} \cdot e^{-
{\tilde{k}^2 \phi^2 \over 24 \Delta^2}} & \qquad \qquad \phi \ll {\Delta
\over \tilde{k}} \nonumber\\
\sum{\cal I}_n = \left( {\tilde{k} \phi \over \Delta} \right)^{\lambda \over
\phi} & \qquad \qquad  {\Delta \over \tilde{k}} \ll \phi < \lambda
\log{\tilde{k} \over \Delta}
\ea
\eea
For larger angles, ladder diagrams are no longer important and the result
will be given by the leading order in perturbation theory. 

For strong coupling $\lambda \gg 1$, $M$ lies in the large $n$ regime. In this 
case, the series is dominated by terms in the large $n$ region, and we find (for small angles)
\bea
\ba{ll}
e^{\sqrt{\lambda} \tilde{k} \over e \Delta} e^{-{1\over 6} \left({\tilde{k} \phi \over \Delta}\right)^2} \le \sum{\cal I}_n \le e^{\sqrt{\lambda} \tilde{k} \over  \Delta} e^{-{1\over 3} \left({\tilde{k} \phi \over \Delta}\right)^2} & \qquad \qquad \phi \ll {\Delta \over \tilde{k}} \nonumber\\
\left({2 \tilde{k} \phi \over \Delta}\right)^{\sqrt{\lambda} \over \phi} \le \sum{\cal I}_n \le \exp\left(\sqrt{2 \lambda \tilde{k} \over \phi \Delta} \right) & \qquad \qquad \phi \gg {\Delta \over \tilde{k}} \ea
\eea

The detailed form of the correlation functions would now be obtained
by 
inserting these expressions for ${\cal I}_n$ into the 
expression (\ref{expr}) and integrating over $\Delta$.  However, the universal 
behavior at large momenta, including the angular dependence, is
already quite clear from the various expressions for ${\cal I}_n$.

 We see that both for the weak coupling and 
the strong coupling expressions, $\sum {\cal I}_n$ is very sharply peaked near 
$\phi=0$, and decays very rapidly for $\phi > {\Delta \over \tilde{k}}$. 
Roughly, the typical value of $\Delta$ in the Fourier transform will be ${1 
\over k}$ so we see that there is a critical angle $\phi = {1 \over k 
\tilde{k}}$ below which the correlation functions are strongly enhanced.\footnote{We should point out that the external momentum dependent phase factor in (\ref{general}) gives rise to an additional angular dependence, $\cos(k \tilde{k} sin \phi)$ which oscillates rapidly for large momenta.} 
For $\phi=0$ we get precisely the exponential dependence on momenta that was observed in the two point functions. 

In \cite{ghi}, it was pointed out that in the absence of a natural normalization of individual operators the most meaningful objects to consider are normalization independent ratios of correlation functions, 
\be
{\langle W_1 (k_1) W_2(k_2) \cdots W_n (k_n) \rangle \over \langle 
W_1 (k_1) W_1 (-k_1) \rangle^{1 \over 2} \cdots \langle W_n 
(k_n) W_n (-k_n) \rangle^{1 \over 2}}
\ee
They argued that the exponential behavior of the two-point functions 
 would generally lead to exponential 
suppression of the correlation functions at high momenta. Given our 
results, we see that this will be true for generic angles, however, 
if the momenta of two operators become pairwise 
antiparallel ($\phi \ll {1 \over k \tilde{k} } $), the numerator will also 
exhibit exponential behavior that should precisely cancel the exponential 
behavior in the denominator. In this case, the momentum dependence is no 
longer universally determined by the Wilson line contribution. Rather
it is determined by the correlation functions of the local operators
attached to the Wilson line. Therefore we expect 
that it reverts to more typical field theoretic behavior. In
particular, in the planar limit the behavior is determined by the UV
fixed point of the commutative version of the theory. It would be very interesting to understand the meaning of the angular dependence we have observed from the point of view of the gravity dual, but we leave this as a question for future work.

\section*{Acknowledgements}

We would like to acknowledge useful conversations and correspondence
with Aki Hashimito, Tom Banks, Micha Berkooz, Steve Giddings, Sunny Itzhaki,
Shamit Kachru, Rob Leigh, Hong Liu, Shiraz Minwalla, Arvind Rajaraman
and Lenny Susskind.

M.R. thanks the theory group at Stanford University for hospitality during the 
initial stages of this work, and the theory group at Harvard
University  for its hospitality in the final stages of the work.
M.V.R. thanks the theory group at Harvard University for hospitality
during part of this work. 
The work of M.R. is supported by DOE grant DOE-FG0296ER40959.
The work of M.V.R is supported in part by the Stanford Institute 
for Theoretical Physics and by NSF grant 9870115.

\end{document}